\documentclass[prb,twocolumn,amssymb,showpacs,amsmath,nobibnotes,superscriptaddress,floatfix,aps]{revtex4}
\usepackage{graphicx}
\usepackage{bm}

\begin{document}
\title {Spectral and optical properties in the antiphase stripe phase of the cuprate superconductors}

\author{Hong-Min Jiang}
\affiliation{National Laboratory of Solid State of Microstructure
and Department of Physics, Nanjing University, Nanjing 210093,
China}
\author{Cui-Ping Chen}
\affiliation{National Laboratory of Solid State of Microstructure
and Department of Physics, Nanjing University, Nanjing 210093,
China}
\author{Jian-Xin Li}
\affiliation{National Laboratory of Solid State of Microstructure
and Department of Physics, Nanjing University, Nanjing 210093,
China}

\date{\today}

\begin{abstract}
We investigate the superconducting order parameter, the spectral
and optical properties in a stripe model with spin (charge)
domain-derived scattering potential $V_{s}$ ($V_{c}$). We show
that the charge domain-derived scattering is less effective than
the spin scattering on the suppression of superconductivity. For
$V_{s}\gg V_{c}$, the spectral weight concentrates on the
($\pi,0$) antinodal region, and a finite energy peak appears in
the optical conductivity with the disappearance of the Drude peak.
But for $V_{s}\approx V_{c}$, the spectral weight concentrates on
the ($\pi/2,\pi/2$) nodal region, and a residual Drude peak exists
in the optical conductivity without the finite energy peak. These
results consistently account for the divergent observations in the
ARPES and optical conductivity experiments in several high-$T_c$
cuprates, and suggest that the "insulating" and "metallic"
properties are intrinsic to the stripe state, depending on the
relative strength of the spin and charge domain-derived scattering
potentials.
\end{abstract}

 \maketitle

\section{introduction}
The nature of spin and/or charge inhomogeneities, especially in the
form of stripes, in some cuprates and their involvement to
high-temperature superconductivity are currently debate
issues.~\cite{kive1} The stripe state is characterized by the
self-organization of the charges and spins in the CuO$_{2}$ planes
in a peculiar manner, where the doped holes are arranged in
one-dimensional (1D) lines and form the so-called "charge stripe"
separating the antiferromagnetic domains. The stripe-ordered state
minimizes the energy of the hole-doped antiferromagnetic system,
thus leading to an inhomogeneous state of matter. Static
one-dimensional charge and spin stripe order have been observed
experimentally in a few special cuprate compounds, specifically in
La$_{1.6-x}$Nd$_{0.4}$Sr$_{x}$CuO$_{4}$~\cite{tran2,zhou1} and
La$_{2-x}$Ba$_{x}$CuO$_{4}$ with $x=1/8$.~\cite{abba1,tran3} Similar
signatures identified in La$_{2-x}$Sr$_{x}$CuO$_{4}$
(LSCO)~\cite{cheo1,maso1,bian1,yama1} and other high temperature
superconductors~\cite{well1,lee4,mook2} point to the possible
existence of stripes, albeit of a dynamical or fluctuating nature.

A pivotal issue about this new electronic state of matter concerns
whether it is compatible with superconductivity, and possibly even
essential for the high transition temperatures, or it competes
with the pairing correlations. A prerequisite for addressing these
issues is to understand the electronic structures of various
stripe states in different cuprates, and to answer the question
whether the stripe phase is intrinsically "metallic" or
"insulating", given its spin- and charge-ordered nature.
Angle-resolved photoemission spectroscopy (ARPES) study by Zhou
\textit{et al.} in (La$_{1.28}$Nd$_{0.6}$Sr$_{0.12}$)CuO$_{4}$
with static stripes have found the depletion of the low-energy
excitation near the ($\pi/2, \pi/2$) nodal region.~\cite{zhou1} In
another compound La$_{1.875}$Ba$_{0.125}$CuO$_{4}$, a system where
the superconductivity is heavily suppressed due to the development
of the static spin and charge orders, Valla \textit{et al.} have
detected the high spectral intensity of the low-energy excitation
in the vicinity of the ($\pi/2, \pi/2$) nodal region [while
antinodal low-energy quasiparticle near $(0, \pi)$ are
gapped].~\cite{vall1} The compound
(La$_{1.4-x}$Nd$_{0.6}$Sr$_{x}$)CuO$_{4}$ ($x=0.10$ and $0.15$)
with static one-dimensional stripe, seems to be an in-between
system, in where the existence of spectral weight around the nodal
region, though weak, has been identified.~\cite{zhou2}

Meanwhile, optical conductivity measurements on the systems with a
stripe phase also display the divergent results. In
La$_{1.275}$Nd$_{0.6}$Sr$_{0.125}$CuO$_{4}$~\cite{dumm1} and
La$_{1.875}$Ba$_{0.125-x}$Sr$_{x}$CuO$_{4}$,~\cite{orto1} a finite
frequency absorption peak with almost disappearance of the Drude
mode in the low-frequency conductivity in several experiments has
been interpreted as collective excitations of charge stripes or as
charge localization from the disorder created by Nd or Ba
substitutions. These observations may support the suggestion that
such stripe-ordered state should be "insulating" in
nature.~\cite{cast1} On the other hand, optical experiment on
La$_{1.875}$Ba$_{0.125}$CuO$_{4}$~\cite{home1} has observed a
residual Drude peak with a loss of the low-energy spectral weight
below the temperature corresponding to the onset of charge stripe
order, which indicates that stripes are compatible with the
so-called nodal-metal
state.~\cite{ando1,zhou3,dumm2,suth1,lee3,home1}

Although, there have been some theoretical studies on the spectral
and optical properties in the stripe phase in the past years
,~\cite{tohy1,mark1,mart1,lore1} the contradictory observations in
recent experiments as mentioned above have yet not been explained
consistently in theoretical frame by adopting a realistic stripe
model. In this paper, by using a stripe model in which the
experimentally observed spin and charge structures at 1/8 doping
are well reflected, we show that the spin domain-derived
scattering will depress the zero-energy spectral weight around the
nodal regions, while the charge domain-derived scattering will
suppress mostly those around the antinodal regions and the hot
spots. Compared to the ARPES data, this suggests that the
different spectral weight distribution may result from the
different relative strength of the spin and charge domain-derived
scattering potentials inherently existing in these compounds.
Meanwhile, a finite frequency peak in the optical conductivity
appears with the disappearance of the Drude peak in the case of
the dominant spin domain-derived scattering. While, when the
charge domain-derived scattering is comparable to the spin one, a
residual Drude peak exists with the disappearance of the finite
energy peak. This suggests that both the "insulating" and
"metallic" properties are intrinsic to the stripe state without
introducing another distinct metallic phase.

The rest of this paper is organized as follows. In Sec. II, we
introduce the model Hamiltonian and carry out the analytical
calculations. In Sec. III, we present the numerical calculations and
discuss the results. In Sec. IV, we present the conclusion.

\section{THEORY AND METHOD}
As the above discussed compounds have a doping density at or near
1/8, we will in this paper consider the 1/8 doping antiphase
vertical stripe state. A schematic illustration of its charge and
spin pattern is presented in Fig. 1. The charge stripes, with a unit
cell of 8 lattice sites (Note for 1/8 doping, there is one hole for
every two sites along the length of a charge stripe), act as
antiphase domain walls for the magnetic order, so that the magnetic
unit cell is twice as long as that for charge order. Due to the
periodical modulation of the stripe order, the electrons moving in
the state will be scattered by the modulation potentials. After
Fourier transformation, the potential $V_{n}$ can be written as the
scattering term between the state $k$ and those at $k\pm nQ$ with
$Q=(3\pi/4,\pi)$. Following Ref.~\onlinecite{mill1}, we expect that
the terms $V_{1}$ and $V_{2}$ will be the dominant spin and charge
domain-derived scattering term, and will be relabeled as $V_{s}$ and
$V_{c}$ in the following, respectively. The weaker higher harmonic
terms will be neglected here. In the coexistence with the
superconducting (SC) order, the model Hamiltonian can be written as
a $16\times16$ matrix for $k$ in the reduced Brillouin zone,
\begin{eqnarray}
\hat{H}=\sum_{k}{'}\hat{C}^{\dag}(k)\left( \begin{array}{cc}
\hat{H}_{k} & \hat{\Delta}_{k}\\
\hat{\Delta}_{k} & -\hat{H}_{k} \\
\end{array} \right)\hat{C}(k),
\end{eqnarray}
where, the prime denotes the summation over the reduced Brillouin
zone. $\hat{C}_{k}$ is a column vector with its elements
$C_{i}(k)=C_{k+(i-1)Q,\uparrow}$ for $i=1,2,\cdots,8$, and
$C^{\dag}_{-k-(i-9)Q,\downarrow}$ for $i=9,10,\cdots,16$. Both
$\hat{H}_{k}$ and $\hat{\Delta}_{k}$  are $8\times8$ matrix with
\begin{widetext}
\begin{eqnarray}
\hat{H}_{k}=\left( \begin{array}{cccccccc}
\varepsilon_{k} & V_{s} & V_{c} & 0 & 0 & 0 & V_{c} & V_{s}\\
V_{s} & \varepsilon_{k+Q} & V_{s} & V_{c} & 0 & 0 & 0 & V_{c} \\
V_{c} & V_{s} & \varepsilon_{k+2Q} & V_{s} & V_{c} & 0 & 0 & 0 \\
0 & V_{c} & V_{s} & \varepsilon_{k+3Q} & V_{s} & V_{c} & 0 & 0 \\
0 & 0 & V_{c} & V_{s} & \varepsilon_{k+4Q} & V_{s} & V_{c} & 0 \\
0 & 0 & 0 & V_{c} & V_{s} & \varepsilon_{k+5Q} & V_{s} & V_{c} \\
V_{c} & 0 & 0 & 0 & V_{c} & V_{s} & \varepsilon_{k+6Q} & V_{s} \\
V_{s} & V_{c} & 0 & 0 & 0 & V_{c} & V_{s} & \varepsilon_{k+7Q}
\end{array} \right),
\end{eqnarray}
\end{widetext}
and
\begin{widetext}
\begin{eqnarray}
\hat{\Delta}_{k}=\left( \begin{array}{cccccccc}
\Delta_{k} & 0 & 0 & 0 & 0 & 0 & 0 & 0 \\
0 & \Delta_{k+Q} & 0 & 0 & 0 & 0 & 0 & 0 \\
0 & 0 & \Delta_{k+2Q} & 0 & 0 & 0 & 0 & 0 \\
0 & 0 & 0 & \Delta_{k+3Q} & 0 & 0 & 0 & 0 \\
0 & 0 & 0 & 0 & \Delta_{k+4Q} & 0 & 0 & 0 \\
0 & 0 & 0 & 0 & 0 & \Delta_{k+5Q} & 0 & 0 \\
0 & 0 & 0 & 0 & 0 & 0 & \Delta_{k+6Q} & 0 \\
0 & 0 & 0 & 0 & 0 & 0 & 0 & \Delta_{k+7Q}
\end{array} \right).
\end{eqnarray}
\end{widetext}
As for the tight-binding energy band, we will choose
the following form,~\cite{lee5,li1}
\begin{eqnarray}
\varepsilon_{k}=&&-2(\delta t+J^{'}\chi_{0})(\cos k_{x}+\cos k_{y})
\nonumber \\&& -4\delta t^{'}\cos k_{x}\cos k_{y}-\mu.
\end{eqnarray}
where, $\delta$ is the doping density, and a $d$-wave SC order
parameter $\Delta_{k}=2J^{'}\Delta_{0}(\cos k_{x}-\cos k_{y})$ is
assumed. Generally, the charge modulation will induce the modulation
of the SC order leading to the finite momentum pairs. However, in
the present study, one of our aim is to examine the effect of the
spin (charge) domain-derived scattering on the SC order. In this
regard, the average value of the SC order parameter is relevant and
the modulation of the SC order will be ignored. We have checked the
effect of this modulation and found no qualitative change in the
results presented in Fig. 2. In the following, $J=100\textmd{meV}$
is taken as the energy unit, $t=2J$, $t^{'}=-0.45t$,
$J^{'}=\frac{3}{8}J$. This dispersion can be derived from the
slave-boson mean-field calculation of the $t-t^{'}-J$
model~\cite{lee5,li1}, and in this way the parameters $\Delta_{0}$,
$\chi_{0}$ and $\mu$ are determined self-consistently. Here we take
it as a phenomenological form. In a self-consistent calculation, the
Hamiltonian is first diagonalized by a unitary matrix $\hat{U}(k)$
with a set of trial values of $\Delta_{0}$, $\chi_{0}$ and $\mu$ for
given potentials $V_{s}$ and $V_{c}$. Then $\Delta_{0}$, $\chi_{0}$
and $\mu$ are self-consistently calculated by using the relations:
$\pm\Delta_{0}=\langle
c_{i\uparrow}c_{i+\tau\downarrow}-c_{i\downarrow}c_{i+\tau\uparrow}\rangle$
(To get the $d$-wave pairing, the sign before $\Delta_{0}$ takes $+$
for $\tau=\pm \hat{x}$ and $-$ for $\tau=\pm \hat{y}$, where
$\hat{x}$ and $\hat{y}$ denote the unit vectors along $x$ and $y$
directions, respectively.), $\chi_{0}=\sum_{\sigma}\langle
c^{\dag}_{i\sigma}c_{j\sigma}\rangle$, and $n=\sum_{\sigma}\langle
c^{\dag}_{i\sigma}c_{i\sigma}\rangle$, respectively.
Reformularization of the expressions of $\Delta_{0}$, $\chi_{0}$ and
$\mu$ in terms of eigenfunctions and eigenvalues of the Hamiltonian,
one obtains the self-consistency relations
\begin{widetext}
\begin{eqnarray}
\Delta_{0}&=&-\frac{1}{N}\sum_{k}(\cos k_{x}-\cos
k_{y})\sum^{16}_{m=1}U_{1m}(k)U^{\dag}_{m9}(k)f[E_{m}(k)]
\nonumber \\
\chi_{0}&=&\frac{1}{N}\sum_{k}(\cos k_{x}+\cos
k_{y})\sum^{16}_{m=1}U_{1m}(k)U^{\dag}_{m1}(k)f[E_{m}(k)]
\nonumber \\
n&=&\frac{2}{N}\sum_{k}\sum^{16}_{m=1}U_{1m}(k)U^{\dag}_{m1}(k)f[E_{m}(k)],
\end{eqnarray}
\end{widetext}
where, $E_{m}(k)$ is the eigenvalue of the Hamiltonian, $U_{mn}(k)$
the elements of the matrix $\hat{U}(k)$, and $f[E_{m}(k)]$ is the
Fermi-Dirac distribution function.

Then, the single particle Green functions
$G_{ij}(k,i\omega_{n})=-\int^{\beta}_{0}d\tau\exp^{i\omega_{n}\tau}\langle
T_{\tau}C_{i}(k,i\tau)C^{\dag}_{j}(k,0)\rangle$ can be expressed
as
\begin{eqnarray}
G_{ij}(k,i\omega_{n})=\sum^{16}_{m=1}\frac{U_{im}(k)U^{\dag}_{mj}(k)}{i\omega_{n}-E_{m}(k)},
\end{eqnarray}
and the spectral functions is
\begin{eqnarray}
A_{ij}(k,\omega)=-\frac{1}{\pi}\textmd{Im} G_{ij}(k,\omega+i0^{+}).
\end{eqnarray}

\section{results and discussion}
\subsection{Self-consistent calculation of the SC order parameter}
We first present in Fig. 2 the self-consistent results of the SC
order parameter as a function of $V_{s}$ and $V_{c}$. While the
scattering from both spin and charge domain-derived scattering
potentials in the stripe state leads to the suppression of the SC
order parameter, the charge domains are more compatible with
superconductivity than spin domains, as can be seen from Fig. 2(a).
This may support the statement that the SC pairing in the stripe
state occurs most strongly within the charge stripes.~\cite{berg1}
On the other hand, an interesting feature is that the SC order
parameter will be zero at the spin domain-derived scattering
potential $V_{sc}\approx 0.14$ in the absence of the charge
domain-derived scattering, however, it will develop a noticeable
value after turning on the charge domain-derived scattering
potential, as shown in Fig. 2(b). This shows that the charge
domain-derived scattering will lead to the emergency of the SC order
which is otherwise destroyed by the spin only scattering.

\subsection{Distribution of spectral weight}
In Fig. 3, we present the distribution of the low-energy spectral
weight in the original Brillouin zone (integrated over an energy
window $\Delta\epsilon=0.1J$ about $\epsilon_{F}$) in the 1/8
antiphase stripe state for different spin (charge) domain-derived
scattering potential $V_{s}$ ($V_{c}$). Let us first look at the
limit where only the spin domain-derived scattering is included,
i.e., $V_{c}=0$ with $V_{s}=0.15$, one will find that the spectral
weight around the nodal region is suppressed heavily[See Fig. 3(a)].
At another limit where only the charge domain-derived scattering is
included ($V_{c}=0.17$ with $V_{s}=0$), the spectral weight around
the nodal region is recovered and those around the hot spot (the
cross of the Fermi surface with the line $k_x\pm k_y=\pm\pi$) and
near the antinodal region are suppressed[See Fig. 3(b)]. Starting
from the limit of $V_{c}=0$ and fixing $V_{s}=0.15$, the spectral
weight will redistribute gradually from the antinodal region to the
nodal region with the increase of the charge domain-derived
scattering potential $V_{c}$, as shown in Figs. 3(c) and (d). When
two scattering potentials are comparable, the strongest spectral
weight situates around the nodal region, and at the meantime
noticeable spectral weights along the whole Fermi surface is
presented. Therefore, the divergent features observed in ARPES
measurements by Zhou \textit{et al.} in
(La$_{1.28}$Nd$_{0.6}$Sr$_{0.15}$)CuO$_{4}$~\cite{zhou2} in which
the low-energy excitations near the nodal region are depleted, and
by Valla \textit{et al.} in
La$_{1.875}$Ba$_{0.125}$CuO$_{4}$~\cite{vall1} in which the high
spectral intensity of the low-energy excitation in the vicinity of
the nodal region is detected are consistently reproduced here by a
change of the relative strength between the charge and spin
domain-derived scatterings. This consistent accounting enables us to
propose that the spin domain-derived scattering dominates over the
charge one in the former system while the scattering strengthes of
them are comparable in the latter system.

In the presence of the spin (charge) domain-derived potential,
quasiparticles near the Fermi surface will be scattered from
$\textbf{k}$ to $\textbf{k}\pm n\textbf{Q}$ (n=1 for the spin
domain-derived potential, n=2 for the charge one), for the 1/8
antiphase vertical stripe configuration shown as Fig. 1. This gives
rise to two scattering channels from the spin domain with potential
$V_{s}$,
\begin{eqnarray}
\textbf{k}&\rightarrow&\textbf{k}+Q=\textbf{k}+(3\pi/4, \pi), \nonumber \\
\textbf{k}&\rightarrow&\textbf{k}-Q=\textbf{k}+(5\pi/4, \pi),
\end{eqnarray}
and two scattering channels from the charge domain with potential
$V_{c}$,
\begin{eqnarray}
\textbf{k}&\rightarrow&\textbf{k}+2Q=\textbf{k}+(3\pi/2, 0), \nonumber \\
\textbf{k}&\rightarrow&\textbf{k}-2Q=\textbf{k}+(\pi/2, 0).
\end{eqnarray}
Strong potential scattering will destruct those parts of the Fermi
surface connected by the above mentioned scattering wave vectors.
Because the scattering wave vectors $Q$ and $-Q$ are close to the
transferred momenta from the node to node scattering, so it will
lead to a depletion of the spectral weight near the nodal region as
shown in Fig. 3(a). On the other hand, the scattering wave vectors
$2Q$ and $-2Q$, which is near the connecting wave vectors between
the two approximately parallel segments of the Fermi surface near
the antinodal and hot spot region, the scatterings with these wave
vectors will suppress the spectral weights around the antinodal and
hot spot regions [Fig. 3(b)].

\subsection{In-plane optical conductivity}
Now, we turn to the discussion of the in-plane optical properties
in the 1/8 antiphase stripe state, and to see how they are
influenced by the scattering from the spin and charge domains. We
will fix the temperature at $T=0.05$ in all calculations, in order
to avoid the influence from the temperature induced change in the
scattering rate. We consider an electric field applied in the $x$
direction, which is perpendicular to the stripe. From the Kubo
formula for the optical conductivity, the real part of the optical
conductivity is $\sigma_{1}(\omega)=-\lim_{q\rightarrow
0}\textmd{Im}[\Pi(q,\omega)]/\omega$. The imaginary part of the
current-current correlation function Im$[\Pi(q\rightarrow
0,\omega)]$ is given by
\begin{eqnarray}
\textmd{Im}[\Pi(q\rightarrow
0,\omega)]=&&\frac{\pi}{N}\sum_{k}{'}\sum^{16}_{j,l=1}v^{jj}(k)v^{ll}(k)
  \nonumber \\&& \times \int d\omega^{'}[f(\omega+\omega^{'})-f(\omega^{'})] \nonumber \\&& \times
 A_{jl}(k,\omega^{'})A_{lj}(k,\omega+\omega^{'}).
\end{eqnarray}
Here, $v^{jj}(k)$ is the diagonal element of the quasiparticle group
velocity in the matrix form
\begin{eqnarray}
\hat{v}(k)=\left( \begin{array}{cc}
\frac{\partial\hat{H}_{k}}{\partial k_{x}} & 0 \\
0 & -\frac{\partial\hat{H}_{k}}{\partial k_{x}} \\
\end{array} \right).
\end{eqnarray}
Figs. 4(a)-4(d) show the results for the optical conductivity
calculated with the same scattering potentials as used to get Fig.
3(a)-3(d). With only spin domain-derived scattering[Fig. 4(a)], no
Drude-like component appears at zero frequency in the optical
conductivity, instead a finite frequency conductivity peak occurs
around 0.3. This indicates that the system exhibits the
"insulating" property.~\cite{note} When only charge domain-derived
scattering is considered[Fig. 4(b)], the Drude-like peak shows up
and at the meantime the finite frequency peak remains. Optical
conductivity involves the contribution from the quasiparticle
excitations along the whole Fermi surface weighted by the
quasiparticle group velocity. Due to the relative flat band
structure near the antinodal region for the high-$T_c$ cuprates,
the zero frequency optical conductivity mainly comes from the
quasiparticle excitations around the nodal region. In the case of
only spin domain-derived scattering, the nodal region of the Fermi
surface is gapped and therefore the quasiparticle spectral weight
is suppressed around the nodal region as shown in Fig. 3(a), so
that the zero-frequency Drude-like peak is absent and a finite
frequency peak with its position being equal to the gap ($\approx
2V_{s}=0.3$) occurs. For the charge domain-derived scattering, the
gap opens around the hot spots and near the antinodal, but a large
spectral weight situates around the nodal region, as can be seen
from Fig. 3(b). Thus, the Drude-like peak emerges and the finite
frequency peak remains (it is now situates at $\approx
2V_{c}=0.34$). As shown in Fig. 3(c), with the increase of the
charge domain-derived scattering $V_{c}$, the gap near the nodal
region which is resulted from the spin domain-derived scattering
will be suppressed gradually and correspondingly the spectral
weight will be enhanced. As a result, the finite frequency peak in
the optical conductivity is shifted to lower frequency and the
zero frequency component is lifted up gradually[Fig. 4(c)]. When
the charge domain-derived scattering is comparable to the spin
one, the quasiparticles have noticeable spectra weight along the
entire Fermi surface with its largest weight around the nodal
region[Fig. 3(d)], then the Drude-like mode occurs at the zero
frequency, and the finite frequency peak fades away and merges
into the Drude-like peak, as shown in Fig. 4(d). The calculated
results for the optical conductivity presented in Figs. 4(c) and
4(d) are consistent well with the experimental observations in the
stripe state of
La$_{1.275}$Nd$_{0.6}$Sr$_{0.125}$CuO$_{4}$~\cite{dumm1} and
La$_{1.875}$Ba$_{0.125}$CuO$_{4}$~\cite{home1}, respectively.

\subsection{Discussion}
We now discuss the implication of our theoretical results. As noted
in the introduction, in La$_{1.275}$Nd$_{0.6}$Sr$_{0.125}$CuO$_{4}$
system, ARPES experiment has found that there is little or no
low-energy spectral weight near the nodal region,~\cite{zhou1} and
optical conductivity experiment has observed a finite frequency peak
with almost the disappearance of the Drude mode, indicating an
"insulating" stripe state.~\cite{dumm1,orto1} These spectroscopic
features can be reproduced here with a strong spin domain-derived
scattering potential $V_{s}=0.15$ and a weak charge domain-derived
potential $V_{c}=0.08$ and $V_{c}=0$, as shown in Figs. 3(a), 3(c),
4(a) and 4(c). Interestingly, in this parameter regime for the spin
and charge domain-derived scattering, the SC order is destroyed as
can be seen from Fig. 2(b). This is in consistent with the
experimental fact that La$_{1.275}$Nd$_{0.6}$Sr$_{0.125}$CuO$_{4}$
is nonsuperconducting. In another cuprate
La$_{1.875}$Ba$_{0.125}$CuO$_{4}$, ARPES spectra have identified the
existence of high spectral intensity around the nodal
region,~\cite{vall1} and the optical conductivity measurement has
observed a residual Drude peak without the finite frequency
peak,~\cite{home1} pointing to a so-called nodal metal
state.~\cite{ando1,zhou3,dumm2,suth1,lee3,home1} When comparable
spin and charge domain-derived scattering potentials are assumed
such as $V_{s}=0.15$ and $V_{c}=0.17$, we can reproduce these
features consistently, as shown in Figs. 3(d) and 4(d). On the other
hand, a weak superconductivity emerges in the otherwise
nonsuperconducting regime (when only spin scattering potential
$V_{sc}$ is considered) with the increase of the charge
domain-derived scattering potential[see Fig. 2(b)]. This suggests
that the weak superconductivity in La$_{1.875}$Ba$_{0.125}$CuO$_{4}$
is likely beneficial from the metallic behaviors of the stripe state
originated from a sufficient charge domain-derived scattering. The
above mentioned consistent accounting for both divergent
spectroscopic features observed in two families of high-$T_c$
cuprates indicates that the stripe state may be intrinsically
"insulating" or "metallic", depending on the relative strength of
the spin and charge domain-derived scattering potentials.
Specifically, a large spin domain-derived scattering potential
favors the "insulating" state, while a large charge domain-derived
scattering potential the "metallic" state.

\vspace*{.4cm}
\section{conclusion}
We have calculated the SC order parameter, the spectral function and
the optical conductivity in a stripe model with spin and charge
domain-derived scattering potentials ($V_{s}$ and $V_{c}$). The
self-consistent calculation of the SC order parameter shows that the
charge domain-derived scattering is less effective than the spin
scattering on the suppression of superconductivity, and may even
lead to the emergency of the SC order which is otherwise destroyed
by the spin only scattering. For $V_{s}\gg V_{c}$, the zero-energy
spectral weight disappears around the nodal points, and a finite
energy peak appears in the optical conductivity with almost the
disappearance of the Drude peak. But for $V_{s}\approx V_{c}$, the
spectral weight concentrates on the nodal region, and a residual
Drude peak exists in the optical conductivity without the finite
energy peak. These results consistently account for the divergent
spectroscopic properties observed experimentally in two families of
high-$T_c$ cuprates, and demonstrate that both the "insulating" and
"metallic" behavior may be the intrinsic properties of the stripe
state, depending on the relative strength of the spin and charge
domain-derived scattering potentials.

\section{acknowledgement}
\par This project was supported by
National Natural Science Foundation of China (Grant No. 10525415),
the Ministry of Science and Technology of Science (Grants Nos.
2006CB601002, 2006CB921800), the China Postdoctoral Science
Foundation (Grant No. 20080441039), and the Jiangsu Planned Projects
for Postdoctoral Research Funds (Grant No. 0801008C).

\vspace*{-.0cm}
\begin{figure}[htb]
\begin{center}
\vspace{-.2cm}
\includegraphics[width=210pt,height=100pt]{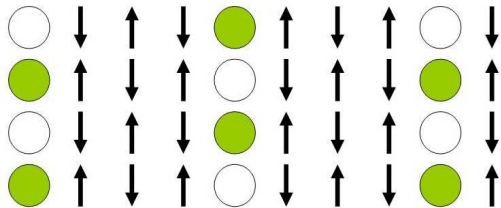}
\caption{(Color online) Schematic illustration of the charge and
spin patterns in the 1/8 doped antiphase stripe state. Circles
represent the charge domain wall (An empty circle indicates a hole
density of one per site), and arrows the copper spins.}\label{fig1}
\end{center}
\end{figure}
\vspace*{-.2cm}
\begin{figure}[htb]
\begin{center}
\vspace{-.2cm}
\includegraphics[width=260pt,height=140pt]{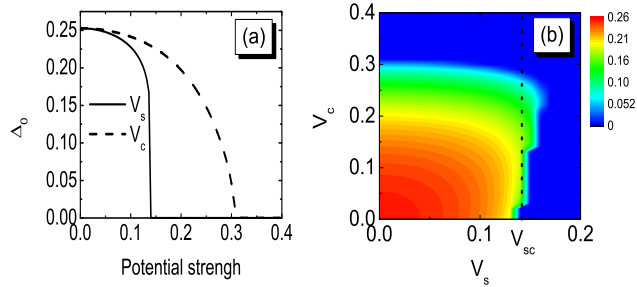}
\caption{(Color online) (a) Superconducting order parameter as a
function of $V_{s}$ and $V_{c}$, respectively. (b) A two-dimensional
map of the superconducting order parameter in the parameter space of
$V_{s}$ and $V_{c}$.}\label{fig2}
\end{center}
\end{figure}
\vspace*{-.2cm}
\begin{figure}[htb]
\begin{center}
\vspace{-.2cm}
\includegraphics[width=260pt,height=230pt]{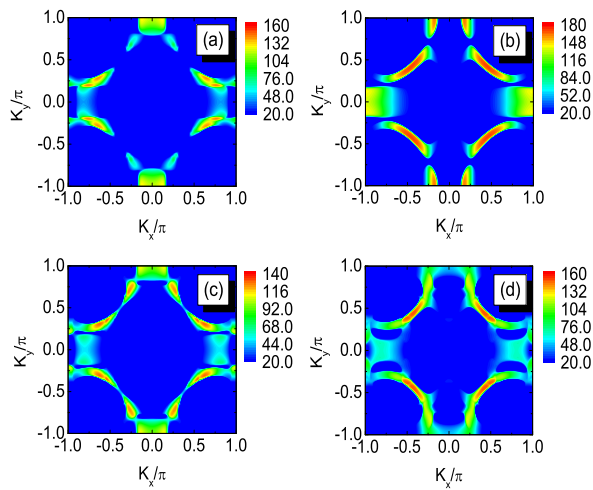}
\caption{(Color online) Spectral weight distribution for different
spin (charge) domain-derived scattering potentials in the normal
state with (a) $V_{s}=0.15$ and $V_{c}=0$, (b) $V_{s}=0$ and
$V_{c}=0.17$, (c) $V_{s}=0.15$ and $V_{c}=0.08$, and (d)
$V_{s}=0.15$ and $V_{c}=0.17$, respectively.}\label{fig3}
\end{center}
\end{figure}
\vspace*{-.2cm}
\begin{figure}[htb]
\begin{center}
\vspace{-.2cm}
\includegraphics[width=240pt,height=210pt]{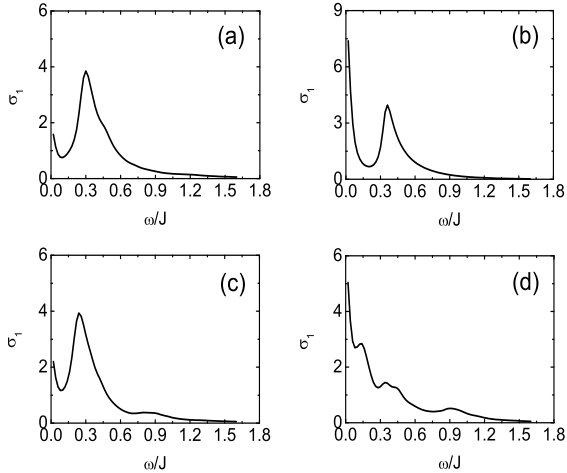}
\caption{In-plane optical conductivity as a function of frequency
for different spin (charge) domain-derived scattering potentials in
the 1/8 antiphase stripe with the SC order parameter $\Delta=0$. (a)
$V_{s}=0.15$ and $V_{c}=0$, (b) $V_{s}=0$ and $V_{c}=0.17$, (c)
$V_{s}=0.15$ and $V_{c}=0.08$, and (d) $V_{s}=0.15$ and
$V_{c}=0.17$.}\label{fig4}
\end{center}
\end{figure}
\vspace*{-.0cm}


\begin{thebibliography}{55}
\bibitem{kive1} S. A. Kivelson and I. P. Bindloss, E. Fradkin, V. Oganesyan, J. M. Tranquada, A. Kapitulnik,
                and C. Howald, Rev. Mod. Phys. \textbf{75}, 1201 (2003).
\bibitem{tran2} J. M. Tranquada, B. J. Sternlieb, J. D. Axe, Y. Nakamura, and S. Uchida,
                Nature \textbf{375}, 561 (1995).
\bibitem{zhou1} X. J. Zhou, P. Bogdanov, S. A. Kellar, T. Noda, H. Eisaki,
             S. Uchida, Z. Hussain, and Z.-X. Shen, Science \textbf{286},
             268 (1999).
\bibitem{tran3} J. M. Tranquada, H. Woo, T. G. Perring, H. Goka, G. D. Gu, G. Xu, M. Fujita, and K. Yamada,
                Nature \textbf{429} 534 (2004).
\bibitem{abba1} P. Abbamonte, A. Rusydi, S. Smadici, G. D. Gu, G. A. Sawatzky and D. L. Feng,
                Nat. Phys. \textbf{1} 155 (2005).
\bibitem{cheo1} S-W. Cheong, G. Aeppli, T. E. Mason, H. Mook, S. M. Hayden, P. C. Canfield, Z. Fisk,
                K. N. Clausen, and J. L. Martinez, Phys. Rev. Lett. \textbf{67}, 1791 (1991).
\bibitem{maso1} T. E. Mason, G. Aeppli, and H. A. Mook, Phys. Rev. Lett. \textbf{68}, 1414 (1992).
\bibitem{bian1} A. Bianconi, N. L. Saini, A. Lanzara, M. Missori, T. Rossetti, H. Oyanagi,
                H. Yamaguchi, K. Oka, and T. Ito, Phys. Rev. Lett. \textbf{76}, 3412 (1996).
\bibitem{yama1} K. Yamada, C. H. Lee, K. Kurahashi, J. Wada, S. Wakimoto, S. Ueki,
                H. Kimura, Y. Endoh, S. Hosoya, G. Shirane, R. J. Birgeneau, M. Greven,
                M. A. Kastner, and Y. J. Kim, Phys. Rev. B \textbf{57}, 6165 (1998).
\bibitem{well1} B. O. Wells, Y. S. Lee, M. A. Kastner, R. J. Christianson, R. J. Birgeneau, K. Yamada,
                Y. Endoh, and G. Shirane, Science \textbf{277}, 1067 (1997).
\bibitem{lee4} Y. S. Lee, R. J. Birgeneau, M. A. Kastner, Y. Endoh, S. Wakimoto, K. Yamada, R. W. Erwin,
               S.-H. Lee, and G. Shirane, Phys. Rev. B \textbf{60}, 3643 (1999).
\bibitem{mook2} H. A. Mook, P. Dai, S. M. Hayden, G. Aeppli, T. G. Perring,
                and F. Do\v{g}an, Nature \textbf{395}, 580 (1998).
\bibitem{vall1} T. Valla, A. V. Fedorov, J. Lee, J. C. Davis, and G. D. Gu, Science \textbf{314}, 1914
               (2006).
\bibitem{zhou2} X. J. Zhou, T. Yoshida, S. A. Kellar, P. V. Bogdanov, E. D. Lu, A. Lanzara, M. Nakamura, T. Noda,
                T. Kakeshita, H. Eisaki, S. Uchida, A. Fujimori, Z. Hussain, and Z.-X. Shen,
                Phys. Rev. Lett. \textbf{86}, 5578 (2001).
\bibitem{dumm1} M. Dumm, and D. N. Basov, Seiki Komiya, Yasushi Abe, and Yoichi Ando,
                Phys. Rev. Lett. \textbf{88}, 147003 (2002).
\bibitem{orto1} M. Ortolani, P. Calvani, S. Lupi, U. Schade, A. Perla, M. Fujita, and K. Yamada,
               Phys. Rev. B \textbf{73}, 184508 (2006).
\bibitem{cast1} C. Castellani, C. Di Castro, M. Grilli, A. Perali, Physica C \textbf{341-348}, 1739 (2000).
\bibitem{home1} C. C. Homes, S. V. Dordevic, G. D. Gu, Q. Li, T. Valla, and J. M. Tranquada,
                Phys. Rev. Lett. \textbf{96}, 257002 (2006).
\bibitem{ando1} Y. Ando, A. N. Lavrov, S. Komiya, K. Segawa, and X. F. Sun,
                Phys. Rev. Lett. \textbf{87}, 017001 (2001).
\bibitem{zhou3} X. J. Zhou, T. Yoshida, A. Lanzara, P. V. Bogdanov, S. A. Kellar,
                K. M. Shen, W. L. Yang, F. Ronning, T. Sasagawa, T. Kakeshita, T. Noda,
                H. Eisaki, S. Uchida, C. T. Lin, F. Zhou, J. W. Xiong, W. X. Ti, Z. X. Zhao,
                A. Fujimori, Z. Hussain, and Z.-X. Shen, Nature (London) \textbf{423}, 398 (2003).
\bibitem{dumm2} M. Dumm, S. Komiya, Y. Ando, and D. N. Basov,
                Phys. Rev. Lett. \textbf{91}, 077004 (2003).
\bibitem{suth1} M. Sutherland, D. G. Hawthorn, R. W. Hill, F. Ronning, S. Wakimoto,
                H. Zhang, C. Proust, E. Boaknin, C. Lupien, L. Taillefer,
                R. Liang, D. A. Bonn, W. N. Hardy, R. Gagnon, N. E. Hussey,
                T. Kimura, M. Nohara, and H. Takagi, Phys. Rev. B \textbf{67}, 174520 (2003).
\bibitem{lee3} Y. S. Lee, K. Segawa, Y. Ando, and D. N. Basov,
               Phys. Rev. B \textbf{70}, 014518 (2004).
\bibitem{tohy1} T. Tohyama, S. Nagai, Y. Shibata, and S. Maekawa, Phys. Rev. Rev. \textbf{82},
              4910 (1999).
\bibitem{mark1} R. S. Markiewicz, Phys. Rev. B \textbf{62}, 1252 (2000).
\bibitem{mart1} I. Martin, G. Ortiz, A. V. Balatsky, and A. R. Bishop,
              Europhys. Lett. \textbf{56}, 849 (2001).
\bibitem{lore1} J. Lorenzana, and G. Seibold, Phys. Rev. Rev. \textbf{90}, 066404 (2003).
\bibitem{mill1} A. J. Millis, and M. R. Norman, Phys. Rev. B \textbf{76}, 220503(R) (2007).
\bibitem{lee5} P. A. Lee, N. Nagaosa, and X. G. Wen, Rev. Mod. Phys. \textbf{78}, 17 (2006).
\bibitem{li1} J. X. Li, C. Y. Mou, and T. K. Lee, Phys. Rev. B \textbf{62}, 640 (2000).
\bibitem{berg1} E. Berg, E. Fradkin, E.-A. Kim, S. A. Kivelson, V. Oganesyan, J. M. Tranquada, and S. C. Zhang,
               Phys. Rev. Lett. \textbf{99}, 127003 (2007).
\bibitem{note} The term "insulating" state used here follows
Refs.~\onlinecite{zhou2,orto1,home1} to indicate a strong
suppression of Drude peak in the optical conductivity. In fact, the
spectral weight at the Fermi level will not be fully gapped out, so
it is not a true insulating state. We use the term here is to
facilitate our comparison with the
experiments[Refs.~\onlinecite{zhou2,orto1,home1}].
\end{thebibliography}
\end{document}